\documentclass[a4paper,fleqn]{cas-dc}

\usepackage{amsmath,amsfonts}
\usepackage{algorithmic}
\usepackage{algorithm}
\usepackage{array}
\usepackage{caption}
\usepackage{textcomp}
\usepackage{subfig}
\usepackage{stfloats}
\usepackage{url}
\usepackage{verbatim}
\usepackage{graphicx}
\usepackage{xcolor}
\usepackage{natbib}
\usepackage[intoc]{nomencl}
\usepackage{tabularx}
\usepackage{adjustbox}
\usepackage{float}
\makenomenclature

\newcommand{\new}[1]{{\color{black}{\textbf{}#1}}}

\begin{document}
\let\WriteBookmarks\relax
\def\floatpagepagefraction{1}
\def\textpagefraction{.001}

\shorttitle{Methods for Mitigating Uncertainty in Real-Time Operations of a Connected Microgrid}

\shortauthors{Panda et~al.}

\title[mode = title]{Methods for Mitigating Uncertainty in Real-Time Operations of a Connected Microgrid}   



 \tnotetext[1]{DesCartes: this research is supported by the National Research Foundation, Prime Minister’s Office, Singapore under its Campus for Research Excellence and Technological Enterprise (CREATE) programme}


%
\author[1]{Subrat Prasad Panda}

 \ead{subratpr001@e.ntu.edu.sg}

\affiliation[1]{organization={NTU}, country={Singapore}}

\author[2]{Blaise Genest}

 \ead{blaise.genest@cnrsatcreate.sg}

\affiliation[2]{organization={CNRS, CNRS@CREATE, IPAL}, country={France \& Singapore}}


\author[1]{Arvind Easwaran}

 \ead{arvinde@ntu.edu.sg}

\author[3]{R\'emy Rigo-Mariani}

\affiliation[3]{organization={Univ. Grenoble Alpes, CNRS, G2Elab, Grenoble}, country={France}}

 \ead{remy.rigo-mariani@g2elab.grenoble-inp.fr}

\author[4]{PengFeng Lin}

\affiliation[4]{organization={EDF R\&D Lab SG}, country={Singapore}}

 \ead{pengfeng.lin@edf.sg}

\begin{abstract}
In this paper, we compare the effectiveness of a two-stage control strategy for the energy management system (EMS) of a grid-connected microgrid under uncertain solar irradiance and load demand using a real-world dataset from an island in Southeast Asia (SEA). The first stage computes a day-ahead commitment for power profile exchanged with the main grid, while the second stage focuses on real-time controls to minimize the system operating cost. Given the challenges in accurately forecasting solar irradiance for a long time horizon, scenario-based stochastic programming (SP) is considered for the first stage. For the second stage, as the most recent weather conditions can be used, several methodologies to handle the uncertainties are investigated, including: 1) the rule-based method historically deployed on EMS, 2) model predictive controller (MPC) using either an explicit forecast or scenario-based stochastic forecast, and 3) Deep Reinforcement Learning (DRL) computing its own implicit forecast through a distribution of costs. Performances of these methodologies are compared in terms of precision with a reference control assuming perfect forecast – i.e. representing the minimal achievable operation cost in theory. Obtained results show that MPC with a stochastic forecast outperforms MPC with a simple deterministic prediction. This suggests that using an explicit forecast, even within a short time window, is challenging. Using weather conditions can, however, be more efficient, as demonstrated by DRL (with implicit forecast), outperforming MPC with stochastic forecast by 1.3\%. 

\end{abstract}


\maketitle


\nomenclature{$T$}{ Sliding window horizon}
\nomenclature{$S_d, S_r$}{\new{Set of scenarios (profiles of PV generation and load demand) used in day-ahead contract and real-time operation respectively}}
\nomenclature{$\mathbb{P}_d(s)$}{Probability of scenario $s \in S_d$}
\nomenclature{$\mathbb{P}_r(s)$}{Probability of scenarios $s \in S_r$}
\nomenclature{$P^t_{pv}$}{PV generation at time $t$ (kW)}
\nomenclature{$P^t_{ld}$}{Load demand at time $t$ (kW)}

\nomenclature{$P^t_{gb}, P^t_{gs}$}{Power buy/sell from the grid at time $t$ (kW)}
\nomenclature{$C^t_{g}$}{\new{Unit cost} of power import/export from the main grid at time $t$ i.e. electricity purchase price (\$/kWh) }
\nomenclature{$P^t_{dESS}, P^t_{cESS}$}{ESS discharge/charge at time $t$ (kW)}
\nomenclature{$E^t$}{ESS SOC at time $t$ (kWh)}
\nomenclature{$R^t_d, R^t_c$}{Down/up reserve for ancillary services at time $t$ (kW)}
\nomenclature{$C_r$}{\new{Unit revenue} for reserve provision (\$/kWh)}
\nomenclature{$\overline{P}_{g}$}{Threshold for main grid power exchange (kW)}
\nomenclature{$u^t_g$}{Binary variable controls main grid exchange buy/sell at time $t$}

\nomenclature{$P^t_{dESS}, P^t_{cESS}$}{ESS discharge/charge at time $t$ (kW)}
\nomenclature{$C_{ESS}$}{\new{Unit cost} of operating ESS (\$/kWh)}
\nomenclature{$\overline{P}_{ ESS }$}{Rated power of ESS (kW)}
\nomenclature{$\eta_d, \eta_c$}{ESS discharge/charge efficiency}
\nomenclature{$E_{end}$}{SOC at the end of sliding window horizon $T$ (kWh)}
\nomenclature{$u^t_{ESS}$}{Binary variable controls ESS charge/discharge at time $t$}
    
\nomenclature{$P^t_{dg}$}{DG power at time $t$ (kW)}
\nomenclature{$DG_{start}$}{SOC value where DG starts (kWh)}
\nomenclature{$DG_{stop}$}{SOC value where DG stops (kWh)}
\nomenclature{$C_{dg}$}{\new{Unit} generation cost of DG (\$/kWh)}
\nomenclature{$RU/RD$}{Up/Down run ramp of DG (kW)}
\nomenclature{$SU/SD$}{Up/Down start-up ramp of DG (kW)}
\nomenclature{$\overline{P}_{dg}$}{DG maximum capacity (kW)}
\nomenclature{$\underline{P}_{dg}$}{DG minimum capacity (kW)}
\nomenclature{$u^t_{dg}$}{Binary variable controls DG status on/off at time $t$}

\printnomenclature

\section{Introduction} \label{introduction}

Given the population growth, energy consumption is estimated to increase by around 80\% from 24700 terawatt hours (TWh) in 2021 to 2050. According to a report by the International Energy Agency, renewable energy sources such as wind and solar accounted for 28\% of electricity generation in 2021 and is expected to increase to 65\% by 2050 \cite{weo2022}. In the context of increasing integration of such distributed energy sources, the concept of microgrids is widely accepted as it promotes the efficient local use of renewable energy by balancing electricity consumption from both renewable and fossil fuel sources. Due to the widespread availability of renewable energy sources, there is a good opportunity to introduce innovative microgrid solutions \cite{irena2018}.

A typical hybrid microgrid then consists of distributed energy sources (DERs) such as renewable energy sources (RESs): photovoltaic (PV), wind turbine (WT), etc.; energy storage systems (ESS) such as lead-acid batteries; diesel generators (DG), etc.
Overall, optimal operations of microgrids depend on the target applications, e.g., cost minimization, maximum use of local generation, commitment to a predicted profile, etc. To supervise and control microgrid operations, Energy Management System (EMS) schedules and manages power flow to/from DERs.

\begin{figure}[b]
    \centering
    \includegraphics[scale=0.50]{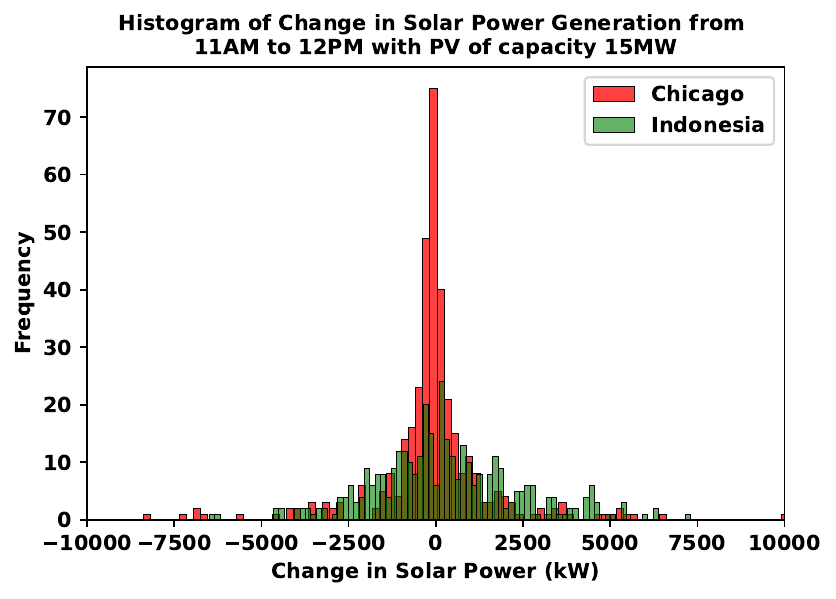}
    \captionsetup{justification=centering}
    \caption{High variability in the solar generation in SEA due to tropical climate.
    }
    \label{fig:sea_vs_us}
\end{figure}

The main challenge in optimizing the operation of microgrids is the mitigation of uncertainties for the demand and local generation, especially due to weather conditions, which is exacerbated in some geographic zones such as SEA \cite{zhao2015, SHOKRIGAZAFROUDI2019201}. As an illustration of solar variability, Figure \ref{fig:sea_vs_us} shows the distribution of differences in solar power generation (in kW) between 11 am and 12 pm over a year for a PV with 15MW capacity. It can be observed that the volatility of solar power generation in Indonesia (SEA) is very high compared to data in Chicago \cite{citylearn}. 

A vast amount of research in the area of microgrid operations focuses on uncertainty mitigation through appropriate scheduling and power allocation of DERs, i.e., deciding when and how much energy to use from specific DERs \cite{SOLEYMANI20081493, SHOKRIGAZAFROUDI2019201}.

EMSs then often rely on forecasts of renewable generation and load demand to optimize DER dispatch, although several research papers propose methods that do not require any forecast \cite{PRATHAPANENI2019100232}. However, a forecast-based approach is applied for real-time operation and optimization in microgrids for its compute efficiency and simplicity \cite{meng2016microgrid, FATHIMA2015431, SHARMA2022120028}. Traditional optimization approaches use forecast-based strategies to formulate the microgrid operation as a mathematical program such as linear program (LP), Mixed Integer Linear Program (MILP), Mixed Integer Non-Linear Program (MINLP), etc. \cite{meng2016microgrid, en11010129, liu2011hybrid, tse2015, tpe2017}. Some related studies also solve this problem using heuristic optimization methods such as genetic algorithm, simulated annealing algorithm, particle swarm optimization and ant colony optimization \cite{tste2010, SHARMA2022120028, FATHIMA2015431, MARZBAND2016265}, and even game theory when several stakeholders are involved \cite{tcdics2017}. 

As already mentioned, one main difficulty in optimizing the operation of (micro) grids \cite{zhao2015, SHOKRIGAZAFROUDI2019201} is the uncertain demand and renewable power generation profiles. To mitigate those uncertainties, studies often employ strategies like Robust optimization or Stochastic Programming, chance-constraint methods, Monte Carlo simulations, etc. \cite{PRATHAPANENI2019100232, AKBARIDIBAVAR2020106425, SHOKRIGAZAFROUDI2019201}.
Furthermore, optimization problems are often split into multiple stages to mitigate uncertainty. The initial stage addresses the optimization problem at a lower granularity (on the order of days), while the subsequent stages handle it at a finer granularity (on the order of seconds) \cite{rigomariani:hal-04124512,8649677,8742669,8945346, GHOLAMZADEHMIR2020102480, Putratama2022}.
Nevertheless, these strategies require the expected values or potential scenarios for solar generation and load demand, which is often challenging to obtain accurately.
Current research focuses on using data-driven, learning-based strategies such as reinforcement learning (RL) to alleviate the need for explicit forecasts. As promising examples, \cite{ji2019real} \cite{NAKABI2021100413} \cite{jpes2022} have applied Deep Reinforcement Learning (DRL) to optimize the economic dispatch in microgrids. \cite{isgte2019, dreem} proposed a similar method for an islanded microgrid. Apart from the application of DRL in EMS, DRL has also been applied to other microgrid challenges, such as load management, demand response planning, and distributed energy management \cite{mocanu2018line, tsg2020, tsg2021, tps2018, zhou2020combined}. \new{While various strategies have been proposed to mitigate uncertainties, determining a suitable strategy for a given situation presents a challenge. This work explores suitable strategies (whether data-driven, MILP, or SP) for specific circumstances.}

This paper considers a simple connected microgrid consisting of PV, ESS, and DG. The operation follows a two-stage approach: (a) day-ahead stage, which schedules the profile for the power exchanged with the main grid based on the predicted production and consumption; and (b) a real-time stage that ensures reliable operation of the microgrid accounting for the actual production and consumption values, and while honouring the day-ahead commitments at the lowest cost.

The main contribution of the paper lies in the implementation and comparison of five different strategies for the real-time operation of the microgrid to mitigate uncertainties on a real-world dataset:  MPC Perfect, MPC Forecast, MPC Stochastic, Rule-based, and DRL, representing broad categories such as forecasting-based, distribution-based, heuristics, and data-driven learning-based approaches which is summarized in Table \ref{tab:intro}. In this work, MPC Perfect uses perfect clairvoyant data rather than a forecast to plan DERs. This will be considered as the theoretical optimum and the reference to assess the performances of the other methods. 

This paper is organized as follows: Section \ref{mgrid_dispatch} introduces the two-stage optimization setup for a grid-connected microgrid for real-time operation. Section \ref{result} analyzes the effectiveness of each strategy in handling uncertainties using a real-world dataset from SEA and provides conclusive remarks on their performance. Subsequently, Section \ref{conclusion} summarizes key findings, acknowledges limitations, and suggests future research directions.

\begin{table*}[]
\caption{Various approaches to solve power scheduling problem in a microgrid.}
\centering
\fontfamily{ptm}\selectfont
    \begin{tabular}{llll}
    \toprule
    \textbf{Type} & \textbf{\begin{tabular}[l]{@{}l@{}}Representative \\ Approach\end{tabular}} & \textbf{Challenges}  & \textbf{Other Algorithms} \\
    \midrule
    Benchmark & MPC Perfect & Impossible to obtain perfect forecast & NA   \\ \hline
    \begin{tabular}[l]{@{}l@{}}Forecasting\\ Based\end{tabular} & MPC Forecast  &  \begin{tabular}[l]{@{}l@{}}Assume Forecast is what is going to happen \\ Need efficient forecast algorithm \\ Need Forecastable weather \end{tabular}  
    & \begin{tabular}[l]{@{}l@{}} Neural Networks based forecasting \\ (\cite{VU20221137})\end{tabular}   \\ \hline
    \begin{tabular}[l]{@{}l@{}}Distribution\\ Based\end{tabular}  & Stochastic &  
    \begin{tabular}[l]{@{}l@{}}A priori knowledge of distribution of weather \\ 
        Does not take into account the latest forecast \end{tabular} & \begin{tabular}[l]{@{}l@{}}Robust optimization \\ (\cite{AKBARIDIBAVAR2020106425}) \\ Chance-constraint method \\ (\cite{PRATHAPANENI2019100232})\\ Monte Carlo simulations \\ (\cite{SHOKRIGAZAFROUDI2019201}) \end{tabular} \\ \hline
    Heuristic & Rule Based & \begin{tabular}[l]{@{}l@{}}Knowledge of domain expert, does not take into \\  account the current forecast or historical data\end{tabular} & \begin{tabular}[l]{@{}l@{}} Greedy, Hierarchical \\ (\cite{tste2010} \\ \cite{SHARMA2022120028} \\ \cite{FATHIMA2015431})\end{tabular} \\ \hline
    \begin{tabular}[l]{@{}l@{}}Data-driven\\ Learning Based\end{tabular} & DRL  &  
    \begin{tabular}[l]{@{}l@{}}
        Amount of data and computational power\\
        Right set of information to take into account\\
        Explainability and lack of guarantees
        \end{tabular}
      & \begin{tabular}[l]{@{}l@{}} Various RL algorithms like \\ DDPG, A2C etc. \end{tabular}  \\
      \bottomrule
    \end{tabular}
    
\label{tab:intro}
\end{table*}

\section{Microgrid Dispatch Formulation} \label{mgrid_dispatch}
\begin{figure*}[!htb]
    \centering
    \subfloat[]{\includegraphics[width=2.2in]{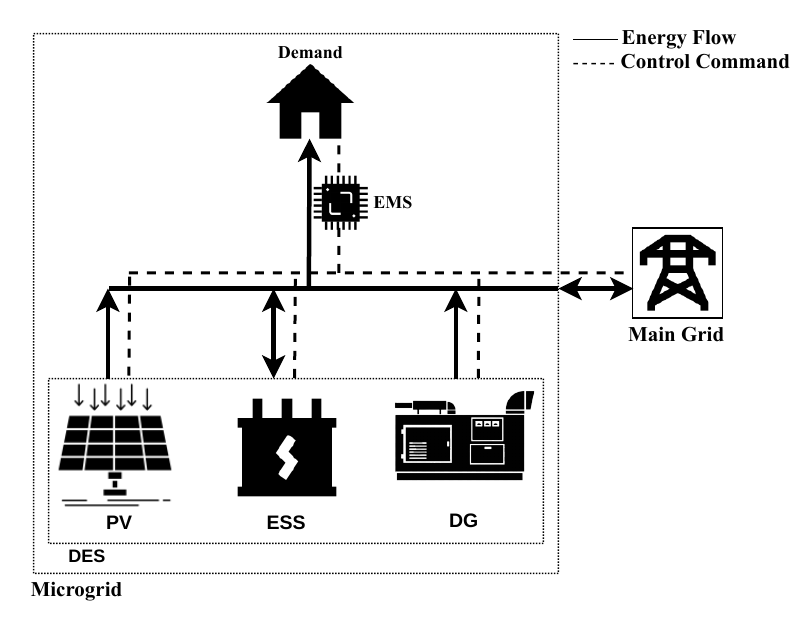}
    \label{fig:arch}}
    \hfil
    \subfloat[]{\includegraphics[width=4.3in]{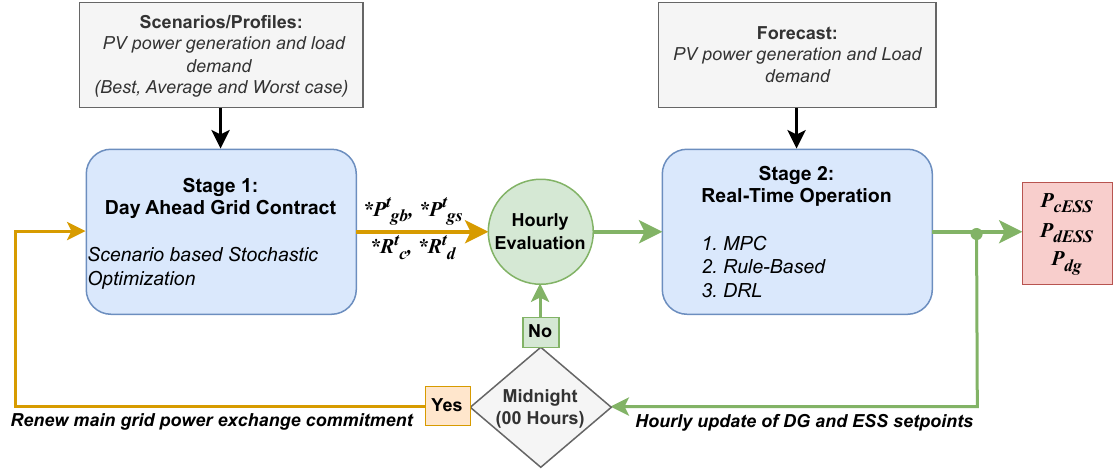}%
    \label{fig:flow}}
    \hfil
    \caption{(a) Grid-connected Microgrid Architecture and (b) A flowchart summarizing the interaction between two stages of the microgrid dispatch problem.}
    \label{fig:microgrid_arch}
\end{figure*}

The microgrid architecture considered in this work is depicted in Figure \ref{fig:arch}. At each time step $t \in [1, 2 \dots T]$ (hourly in our case, where $T$ represents the length of the look-ahead horizon - e.g. 24 Hrs), the load demand $P_{ld}^t$ can be met through various sources, which include: power generated from PV, denoted as $P_{pv}^t$; power from the DG, denoted as $P_{dg}^t$; power imported (bought) from the main grid, denoted as $P_{gb}^t$; and power discharged from the ESS, denoted as $P_{dESS}^t$. The state of charge (SOC) of the ESS, represented as $E^t$, indicates the amount of energy available for discharge from the ESS. Also, any surplus, if needed, can be stored in the ESS by charging it, indicated as $P_{cESS}^t$, or exported (sold) to the main grid, represented as $P_{gs}^t$. We also consider the provision to reserve ESS charge/discharge power for ancillary services, specifically for frequency regulation services to maintain frequency stability. Thus, a part of ESS power is reserved for upward regulation (by discharging ESS) is denoted as $R_d^t$, while the ESS power reserved for downward regulation (by charging ESS) is represented as $R_c^t$. 

\new{The operation cost of each power source depends linearly on the amount of power generated/utilized through a predefined unit cost (cost per unit of power). Here, the unit cost of importing power from the main grid at time $t$ is denoted as $C_{g}^t$, i.e., the electricity purchase price (varies based on the time of the day, so dependent on $t$). We gain revenue by exporting to the main grid accordingly. The unit cost of operating the ESS, homogeneous to the levelized storage cost, is $C_{ESS}$. Similarly, the unit cost of generating power from DG is $C_{dg}$. The unit revenue from reserve provision for ancillary services is $C_r$.} The role of the EMS in the microgrid is to determine the optimal power setpoint for each power source (PV, DG, ESS, and main grid) at each time step $t$ to minimize the overall operation cost. We need to determine the power setpoint of various power sources, particularly values of seven parameters $P^t_{gb}$, $P^t_{gs}$, $R^t_d$, $R^t_c$, $P_{cESS}^t$, $P_{dESS}^t$, and $P_{dg}^t$, based on the generated power from PV $P_{pv}^t$ and to meet the load demand $P_{ld}^t$. Equation \ref{eq:obj_mgrid} represents the primary objective of the microgrid in this work, which is to minimize the operation cost associated with power utilization from various sources over a horizon of length $T$, while ensuring that the load demand can be met.

\begin{multline} \label{eq:obj_mgrid}
  \min \sum_{t=1}^{T}{ \Biggl[ C_{g}^t (P_{gb}^t -  P_{gs}^t) - C_r (R_d^t +R_c^t}) \\
  + C_{ESS} (P_{dESS}^t + P_{cESS}^t) + C_{dg} P_{dg}^t  \Biggr]
\end{multline}

Solving the optimization problem presented in Equation \ref{eq:obj_mgrid} poses the following challenges: (a) the main grid requires power exchange and ESS reserve commitments to be made a day ahead, which the microgrid must strictly adhere to --- the EMS needs to determine four optimal power setpoints ($P_{gb}^t$, $P_{gs}^t$, $R_d^t$, and $R_c^t$) a day in advance and honour these commitments in the following day, and (b) forecasting PV power generation and load demand is difficult for a longer horizon \cite{pvsc2020}.

Solving for all seven parameters a day ahead is inefficient, as it would introduce uncertainty due to forecast errors. Instead, we break down the problem into two stages, allowing us to adjust optimal power setpoints of DG ($P_{dg}^t$) and ESS ($P_{cESS}^t$ and $P_{dESS}^t$) to tackle uncertainties related to PV power generation and load demand in real-time, with the remaining power setpoints fixed due to the commitment made in the first stage. In essence, we split the optimization problem in Equation \ref{eq:obj_mgrid} into two separate optimization problems which is also illustrated in Figure \ref{fig:flow}: 
\begin{enumerate}
    \item Day-ahead Planning: In the first stage, day-ahead contracts are generated by using multi-scenario Stochastic Programming to determine the optimal schedules for $P_{gb}^t$, $P_{gs}^t$, $R_d^t$, and $R_c^t$ --- i.e. grid power exchanges and ESS reserve. This identifies optimal setpoints for the day-ahead commitment that are suitable under different potential scenarios, considering various profiles of PV power generation and load demand (such as best, average, and worst-case scenarios). 
    
    \item Real-Time Operation: In the second stage, the problem for real-time operation is formulated, where power dispatch setpoints are required to be assigned to DG ($P_{dg}^t$) and ESS ($P_{cESS}^t$ and $P_{dESS}^t$) in actual operation, strictly adhering to the day-ahead power exchanges commitments. 
\end{enumerate}

\new{
We use the following widely used constraints in the literature on grid-connected microgrid architecture \cite{7355371, JIN2017480, DIPIAZZA20171}:

(a) power balance: at any time step $t$, the combined power from PV ($P^t_{pv}$), DG ($P^t_{dg}$), main-grid import ($P^t_{gb} - P^t_{gs}$), and ESS ($P^t_{dESS} - P^t_{cESS}$) must exceed the load demand $P^t_{ld}$, allowing for surplus energy to be curtailed as shown in Equation \ref{eq:power_balance}. 
\begin{equation}
    P^t_{pv} + P^t_{gb} - P^t_{gs} + P^t_{dESS} - P^t_{cESS} + P^t_{dg} \geq P^t_{ld} \quad \forall t \in T
    \label{eq:power_balance}
\end{equation}

(b) grid constraint: the microgrid can only buy or sell energy from the main grid (not both), which is decided with the binary variable $u^t_{g}$. A predefined threshold $\overline{P}_g$ bounds grid exchange power as shown in Equation \ref{eq:grid_constraint}. 
\begin{equation}
    \begin{aligned} 
    0 &\leq P^t_{gb} \leq u^t_g \times \overline{P}_g & \quad \forall t \in T  \\
    0 &\leq P^t_{gs} \leq (1 - u^t_g) \times \overline{P}_g & \quad \forall t \in T
    \label{eq:grid_constraint}
\end{aligned}
\end{equation}

(c) ESS charge/discharge constraint: ESS can either be charged or discharged, and the SOC of the battery is updated based on the magnitude of the charge and discharge power. Also, the charge/discharge power of ESS, including reserve power, should not exceed the predefined rated power of ESS denoted by $\overline{P}_{ ESS }$. The binary variable $u^t_{ ESS }$ denotes the charging mode (charging or discharging), and Equation \ref{eq:ess_constraint_soc} updates the SOC represented by the energy within the battery $E^t$, where $\eta_d$ and $\eta_c$ represent the efficiency of ESS discharge and charge, respectively. Equation \ref{eq:ess_constraint_end} ensures that the SOC at the end of the sliding window horizon equals to the predefined parameter $E_{end}$.
\begin{equation}
    \begin{aligned} \label{eq:ess_constraint_bound}
       0 &\leq P_{dESS} \leq u^t_{ESS} \times \overline{P}_{ESS} & \quad \forall t \in T \\
       0 &\leq P_{cESS} \leq (1 - u^t_{ESS}) \times \overline{P}_{ESS} & \quad \forall t \in T
    \end{aligned}
\end{equation}
\begin{equation} \label{eq:ess_constraint_soc}
   E^t = E^{t-1} - \eta_d P^t_{dESS} + \eta_c P^t_{cESS} \quad \forall t \in T
\end{equation}
\begin{equation} \label{eq:ess_constraint_end}
   E^T \geq E_{end}
\end{equation}

(d) ancillary services constraint: we reserve a certain amount of ESS charge/discharge power for upward/downward regulation, limited to the rated power capacity of ESS $\overline{P}_{ ESS }$ given in Equation \ref{eq:reserve_constraint_bound}. Equation \ref{eq:reserve_constraint_pow} ensures that the sum of the scheduled charge/discharge power ($P^t_{dESS}, P^t_{cESS}$) and the reserve power ($R^t_d$, $R^t_c$) does not exceed the rated power of the ESS $\overline{P}_{ ESS }$. 
\begin{equation}
   \begin{aligned} \label{eq:reserve_constraint_bound}
    R^t_d &\leq (1 - u^t_g) \times \overline{P}_{ESS} & \quad \forall t \in T \\
    R^t_c &\leq u^t_g \times \overline{P}_{ESS} & \quad \forall t \in T
\end{aligned} 
\end{equation}
\begin{equation}
    \begin{aligned} \label{eq:reserve_constraint_pow}
    R^t_d + P^t_{dESS} - P^t_{cESS} &\leq \overline{P}_{ESS} & \quad \forall t \in T \\
    R^t_c - P^t_{dESS} + P^t_{cESS} &\leq \overline{P}_{ESS} & \quad \forall t \in T
\end{aligned}
\end{equation}

(e) DG constraint: it ensures start-up and shutdown cannot occur simultaneously, and the power variation of the DG power over two successive time steps remains within the up/down ramp rate bound. $\alpha^t_{dg}$ and $\beta^t_{dg}$ are the start and stop commands for DG. $u^t_{dg}$ is a binary variable that
represents the on/off status of the DG. Equations \ref{eq:dg_constraint_comm} and \ref{eq:dg_constraint_status} show the relationship between the start/stop command for the DG and the DG status at time $t$ --- i.e. start-up and shutdown cannot occur simultaneously, and a logical relationship between the on/off status and the start/stop variables. Equation \ref{eq:dg_constraint_bound} sets a bound on the DG power by the predefined maximum capacity $\overline{P}_{dg}$ and minimum capacity $\underline{P}_{dg}$. Equation \ref{eq:dg_constraint_ramp} ensures that the power variation of the DG power over two successive time steps remains within the up run-ramp $RU$ and the down run-ramp $RD$ limits. In addition, the power variation is limited to a start-up-ramp $SU$ and a shutdown-ramp $SD$ when starting and stopping the generator.
\begin{equation}\label{eq:dg_constraint_comm}
    \alpha^t_{dg} + \beta^t_{dg} \leq 1 \quad \forall t \in T
\end{equation}
\begin{equation} \label{eq:dg_constraint_status}
    \alpha^t_{dg} - \beta^t_{dg} \leq u^t_{dg} - u^{t-1}_{dg} \quad \forall t \in T
\end{equation}
\begin{equation}
    \begin{aligned} \label{eq:dg_constraint_bound}
    P^t_{dg} &\leq u^t_{dg} \times \overline{P}_{dg} & \quad \forall t \in T \\
    P^t_{dg} &\geq u^t_{dg} \times \underline{P}_{dg} & \quad \forall t \in T
    \end{aligned} 
\end{equation}
\begin{equation}
   \begin{aligned} \label{eq:dg_constraint_ramp}
    P^t_{dg} - P^{t-1}_{dg} \leq u^{t-1}_{dg} \times RU + \alpha^t_{dg} \times SU & \quad \forall t \in T \\
    P^{t-1}_{dg} - P^t_{dg} \leq u^{t}_{dg} \times RD + \beta^t_{dg} \times SD & \quad \forall t \in T
\end{aligned} 
\end{equation}

}
The subsequent section is structured as follows: Subsection \ref{model:contract} elaborates on generating the day-ahead grid contract. Sections \ref{model:mpc} through \ref{model:rl} discuss different approaches used for real-time operations, namely Rule-based, MPC, and DRL.

\subsection{Day Ahead Grid Commitments} \label{model:contract}

To generate the day-ahead grid commitment in the first stage, our objective is to determine the optimal setpoint for grid power exchange and ESS ancillary reserve a day in advance. This is achieved using the multi-scenario Stochastic Programming (SP) approach, which aims to identify the schedules of main grid exchange ($P^t_{gb}$, $P^t_{gs}$) and ESS reserve ($R^t_d$, $R^t_c$) considering various scenarios of PV generation and load demand profiles. We consider a set of scenarios $S_d$, such as maximum, average and minimum, where each scenario $s \in S_d$ contains profiles on PV power generation and load demand along $T$ time steps. For instance, the maximum profile considers the maximum possible PV generation or load demand at any particular time step. Similarly, the average profile considers average PV generation/load demand at any given time. We assign a probability of occurrence $\mathbb{P}_d(s)$ to each scenario. 

The objective in Equation \ref{eq:day_ahead} allows finding the suitable commitment that optimally performs across all scenarios, considering their associated probabilities and constraints previously introduced. In this stage, the remaining parameters DG power ($^s\!P_{dg}^t$) and ESS power ($^s\!P_{cESS}^t$, $^s\!P_{dESS}^t$) for each scenario $s \in S_d$ can take any values that are compatible with microgrid constraints described previously.

\begin{equation} \label{eq:day_ahead}
\begin{split}
  \min_{\substack{P^t_{gb}, P^t_{gs} \\ R^t_d, R^t_c}} \sum_{t=1}^{T}{ \Biggl[ C_{g}^t ( P_{gb}^t - P_{gs}^t ) - C_{r} (R_{d}^t + R_{c}^t}) +  \\
\sum_{s \in S_d}{ \mathbb{P}_d(s) \biggl( C_{ESS} ( ^s\!P_{dESS}^t + \ ^s\!P_{cESS}^t) + C_{dg}\ ^s\!P_{dg}^t  \biggr) \Biggr] } 
\end{split}
\end{equation}

We represent the final commitment with the main grid in the first stage, which the second stage (real-time operation) needs to follow strictly, i.e., the solution of Equation \ref{eq:day_ahead} as, $^*\!P_{gb}^t$, $^*\!P_{gs}^t$, $^*\!R_d^t$, and $^*\!R_c^t$. The remaining energy requirements in the microgrid should be met from DG or ESS, which must be decided in real-time operation.

\subsection{Rule-Based Real-Time Operation}\label{model:rule}

\begin{figure}[!htb]
    \centering
    \includegraphics[scale=0.85]{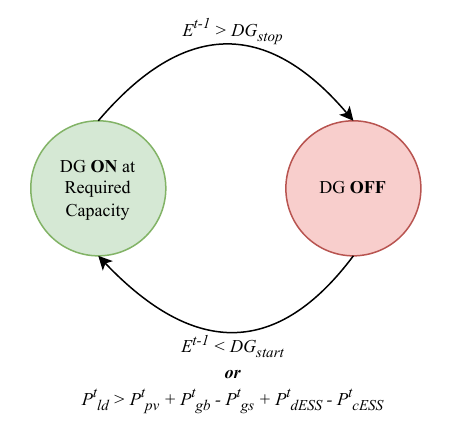}
    \captionsetup{justification=centering}
    \caption{Rule-based real-time operation.}
    \label{fig:rule_based}
\end{figure}

For easy integration and low computationally intensive solutions, the way to control real-time operations can rely on a set of rules. Actions are taken at every time step based on actual measurements. For instance, a cycle charge rule-based algorithm for the real-time operation as in \cite{dreem} can be implemented to determine the start or stop sequences of the generator based on the ESS SOC observed levels – i.e. setpoints of $DG_{start}$ and $DG_{stop}$ respectively. Figure \ref{fig:rule_based} shows the state transition diagram of the rule-based operations considered in this paper. $DG_{start}$ denotes battery SOC $E^t$ at which the generator should start and $DG_{stop}$ denotes battery SOC at which the generator should stop. Whenever the generator starts, it delivers a power set by default at the optimal capacity, which is 75\% of its nominal capacity. However, the generator setpoint is curtailed or increased according to the real-time power requirement.


\subsection{MPC for Real-Time Operation} \label{model:mpc}

Another option for real-time operation is to adopt an MPC approach with a rolling horizon - i.e. look-ahead predictions are updated successively along the operating horizon. The objective is to strictly adhere to the day-ahead commitment with the main grid and minimize the operation cost, i.e., determining the optimal power profiles. MPC adjusts the ESS charge/discharge and the DG power to ensure cost-efficient operation and mitigate uncertainties. The solution obtained for ESS charge/discharge $P^1_{dESS}$, $P^1_{cESS}$, and DG power $P^1_{dg}$ at $t=1$ (current time step) is used as the set-point for real-time operation, and solutions of other time steps are discarded, as is typically done in MPC. The objective for real-time operation follows Equation \ref{eq:obj_mgrid}, where $P^t_{gb}, P^t_{gs}, R^t_d, R^t_c$ are the constants computed in the first stage.


To forecast solar generation and load demand, we use (a) forecasting strategy based on exponential averaging discussed in subsection \ref{model:forecast} and (b) multi-scenario stochastic programming, which considers multiple scenarios to forecast as discussed in subsection \ref{model:sp}

\subsubsection{MPC with Exponential Averaging Forecast} \label{model:forecast}
To assess the performance of the proposed control strategy, forecasts for demand and solar generation are first emulated using a classical approach called exponential moving averaging (EMA) \cite{ema}, where recent data are more important than older ones to predict the next values of time series.
The EMA formulation is provided in Equation \ref{eq:ema}, where $y^{t}$ denotes the moving average at present time instance, $x^{t}$ denotes the present value and $\theta$ denotes the smoothing constant: bigger the $\theta$, the more historical data are taken into account. Here, $y^{t+1}$ is the forecasted value for the next step.

\begin{equation}\label{eq:ema}
    y^{t+1} = \theta \times y^{t} + (1-\theta) \times x^{t}
\end{equation}

To account for the fact that time of day (8am, midnight, etc.) plays a crucial role in both the load and the solar forecast,
$t$ is set to be the day, and the data $y^t$ are those of the previous days at the same time point (we keep 24 rolling averages $y^t_h$, one for each time of day $h \in {0, \ldots, 23}$).
For instance, to forecast solar/demand power at 8 AM tomorrow, historical data of solar generation/load demand only at 8 AM until yesterday is used to compute $y^t_8$.

Further, to take into account {\em local effect} (rainy day, sunny day, etc.), we introduce a scaling factor denoted as $sf^{t+1}$. It helps to capture information about how well the EMA's prediction performed over the past. Here, we compute the scaling factor based on the last few predictions $y^{t-2}$, $y^{t-1}$, and $y^t$
and their ratio with the actual values $x^{t-2}$, $x^{t-1}$, and $x^t$. Equation \ref{eq:scale} is used to compute the scaling factor where $\kappa$ is a constant weighting factor to give more preference to more recent local data. The forecast is obtained by multiplying the EMA forecasted solar generation/load demand value with the scaling factor $sf^{t+1}$.

\begin{equation}\label{eq:scale}
    sf^{t+1} = \frac{1}{\sum_{j=0}^{j=2} (\kappa)^{j+1}} \sum_{j=0}^{j=2} 
    \Biggl[
    (\kappa)^{j+1} \times
    \begin{cases}
        \frac{ x^{t-j}}{y^{t-j} },& \text{if }  y^{t-j} > 0 \\
         1,              & \text{otherwise}
    \end{cases}
    \Biggl]
\end{equation}

\subsubsection{MPC with Stochastic Forecast} \label{model:sp}
Similar to the day-ahead contract generation, a scenario-based Stochastic Programming can be used for the real-time operation, which is shown to handle uncertainties in multiple works of literature \cite{AKBARIDIBAVAR2020106425, SHOKRIGAZAFROUDI2019201}. We fix the first stage's grid exchange commitment and ESS reserve setpoints. 

As shown in Equation \ref{eq:obj_real_time_sp}, multiple scenarios, denoted by $s \in S_r$, which include maximum, average and minimum PV generation/load demand scenarios along with two additional scenarios (one between maximum and average, and the other between average and minimum PV generation/load demand). Each scenario has a probability of occurrence $\mathbb{P}_r(s)$. These scenarios include multiple PV power generation and demand load profiles. Using these scenarios, we find the optimal setpoints for the DG ($P_{dg}^1$) and ESS ($P_{cESS}^1$, $P_{dESS}^1$) at the current time step (at $t=1$), while allowing the power setpoints for the next time steps ($t \neq 1$) to have any values that satisfy microgrid constraints. This approach determines a robust setpoint for the current time step, capable of accommodating multiple possible scenarios. The rolling horizon approach is used to find the optimal setpoint for the subsequent time steps.

\begin{equation} \label{eq:obj_real_time_sp}
\begin{split}
    \min_{\substack{P^{1}_{dg} \\ P^{1}_{cESS}, P^{1}_{dESS}}}
  \sum_{s \in S_r} \mathbb{P}_r(s)
  \sum_{t=1}^{T}{ \Biggl[ C_{g}^t\ ( ^*\!P_{gb}^{t} - ^*\!P^{t}_{gs} ) - C_{r}(^*\!R_{d}^{t} +\ ^*\!R_{c}^{t}) } \\
  + C_{ESS}( ^s\!P_{dESS}^{t} +\  ^s\!P_{cESS}^{t}) + C_{dg} \ ^s\!P_{dg}^{t}  \Biggr] 
\end{split}
\end{equation}

\subsection{DRL for Real-Time Operation}\label{model:rl}

Another way to mitigate the uncertainties is to train a real-time controller using Reinforcement Learning. Here, a Markov Decision Process (MDP) is implemented while considering tuples of "States", "Actions", and "Reward" $(S, A, R)$  \cite{sutton2018reinforcement}.

The set state (observations) $S$ is represented as a six-dimensional vector $\big< h, P^h_{ld}, P^h_{pv}, E^{h}, E^{0}, P^{h-1}_{dg} \big>$, where $h$ is the hour of the day (integer value). $P^h_{ld}$ and $P^h_{pv}$ denote measured load demand and solar generation at hour $h$ respectively. $E^{h}$ is SOC of ESS at the hour $h$ and $E^{0}$ is SOC of ESS at the beginning of the day i.e. at 00 Hrs of the day in 24 Hrs format. The day-ahead commitment is done at the beginning of each day, so the SOC at the beginning of the day $E^{0}$ is included in the state, providing information about the day (such as commitment with the main grid) to the DRL agent. $P^{h-1}_{dg}$ is the generator power at the previous hour. The action set $A$ is a set of finite actions denoting the generator power at the hour $h$. Since we strictly adhere to the contract, energy buying/selling from the main grid and ESS reserves are fixed in the first stage. The remaining power requirements should be met by either DG or ESS. The DRL agent determines DG setpoint $P^t_{dg}$, and then the ESS setpoint ($P^t_{cESS}$ or $P^t_{dESS}$) is determined by subtracting the remaining power requirements from the DG setpoint following the power balance constraint in microgrid. The reward function $R$ is the immediate reward received when the agent takes action $a_s \in A$ from a state $s \in S$, shown in Eq. \ref{reward}. In the considered implementation, the objective is to avoid the condition that the demand is not met (blackouts) while reducing the cost of operation $C(s, a_s)$ with a weighting factor of $w_b$ and $w_o$, respectively. 

\begin{equation} \label{reward}
  R(s, a_s) = -w_o \times C(s,a_s) -w_b \times 
    \begin{cases}
        1 &\text{for blackout}\\
        0 &\text{otherwise}
    \end{cases}
\end{equation}

Since the real-time actions are computed with an MDP, we use DQN (Deep-Q Learning) \cite{mnih2015human} to train the agent to perform appropriate actions to maximize the total reward. DQN has been shown to outperform other DRL approaches for the microgrid applications related to the use case considered in our work \cite{dreem}. The DRL agent (i.e. the controller) then generates an hourly setpoint of DG power for real-time operation, given the system's current state. Once the system is controlled along the simulation horizon, the performances of the proposed approach can be assessed.
\section{ Experiments and Analysis of Results}\label{result}

The experiments were performed on a machine with AMD Ryzen Threadripper PRO 3975WX 32-Cores, 128GB RAM, and NVIDIA GeForce RTX 3090 24GB. The available dataset consists of hourly load demand and PV generation for 2019, which EDF R \& D Lab, Singapore, provides out of one their project in Indonesia. All programs are written in Python, the software library Stable-Baseline3 \cite{stable-baselines3} is used to train RL agents and the Gurobi solver is used as the MILP solver for MPC stages. The average cost of microgrid operation, generator power usage, solar energy consumption and SOC of ESS for the month of December is used as the test set for performance comparison among the methods investigated for real-time operations: (a) MPC Perfect (for benchmarking), (b) MPC Forecast as in \cite{7370817, tcst2014}, (c) MPC Stochastic as in \cite{AKBARIDIBAVAR2020106425, SHOKRIGAZAFROUDI2019201} (d) Rule-based as in \cite{dreem} and (e) DRL. 

\begin{figure}[!htb]
    \centering
    \includegraphics[scale=0.55]{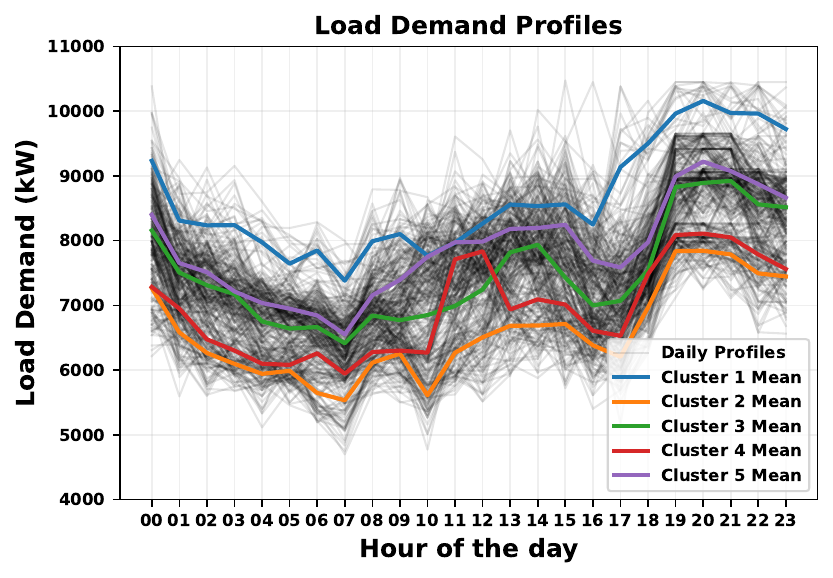}
    \captionsetup{justification=centering}
    \caption{Load demand profile and five cluster heads of the dataset.}
    \label{fig:demand_profile}
\end{figure}

\begin{figure}[!htb]
    \centering
    \includegraphics[scale=0.55]{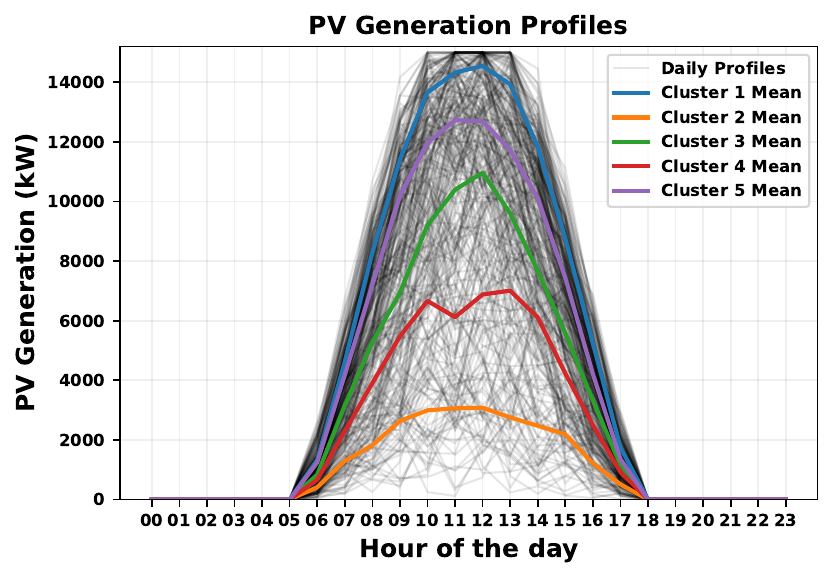}
    \captionsetup{justification=centering}
    \caption{PV generation profile and five cluster heads of the dataset.}
    \label{fig:solar_profile}
\end{figure}

\subsection{Dataset}

The dataset contains hourly entries of load demand and PV power generation for a year. It involves high variance in samples, for instance, sudden cloud coverage during data sampling can result in lower recorded values for PV power generation than their actual values. Data with high variance is valuable for testing the robustness of optimization strategies. The load demand in the dataset ranges from \new{5000 kW to 10000 kW}, while PV power generation ranges from \new{0 to 15000 kW.}  

Figures \ref{fig:demand_profile} and \ref{fig:solar_profile} illustrate the set of daily profiles of load demand and PV power generation for the year. Each profile represents a specific day's hourly load demand or PV generation. The black line in the plot represents the profile for each day in the dataset. In the experiment, all profiles from the entire year are clustered into five different clusters using K-Means clustering, and the mean of each cluster (cluster head) is used in MPC Stochastic. Note that, from the plots, it is obvious that the variance in load demand is much lower compared to the variance in PV power generation. This suggests that weather conditions have a significant role in contributing to uncertainties in microgrid operation.

\subsection{Experiment Setup}
The dataset is divided into two parts: one for training and one for testing. The training data comprises data from January through November, while the testing data only includes data from December. The training data is utilized in different ways by various approaches. It is incorporated into the forecasting algorithm for MPC Forecast to predict PV generation and load demand for future time horizons. In contrast, in MPC Stochastic, it is used to produce solar generation and load demand profiles by clustering the data as shown in Figures \ref{fig:demand_profile} and \ref{fig:solar_profile}. Lastly, it is also used to train the agent in Deep Reinforcement Learning (DRL). 

Notice that the testing is conducted for the entire month of December, with all approaches starting on the first of December and continuing with hourly data until the end of the month, without resetting the algorithm in between. Hence, some methods can outperform MPC Perfect on certain days, for example, when the difference between the SOC at the end and the beginning of the day is higher than that of MPC Perfect (this difference is adjusted in the previous or subsequent days)

Table \ref{tab:parameters} shows the parameters used for the experiment and \new{Table \ref{tab:parameters2} provides the unit cost of power exchange $C^t_g$ with the main grid for different hours of the day.} A rolling horizon of 24 Hrs is considered for MPC Perfect, MPC Forecast and MPC Stochastic is 24 hours. All the approaches for real-time operation have an operational granularity of one hour, i.e. the setpoints for DERs are computed at one-hour intervals. In the case of the DQN agent, a network with four hidden layers containing 64, 128, 128, and 64 neurons, respectively, is suitable for real-time operation. The action space consists of 40 discrete actions. During the training of the DQN agent, a discount factor of 0.9 is used, and the learning rate is set to 0.001.

\begin{table}[]
\caption{Values of different parameters for the experiment}
\centering
\fontfamily{ptm}\selectfont
\begin{tabular}{ll}
\toprule
Parameters & Value Assumed \\ 
\midrule
$T$ & 24 \\ 
$|S_d|$ & 3 \\
$|S_r|$ & 5 \\ 

$\overline{P}_g$ & 5000 kW \\

$\overline{E}$ & 25000 kWh \\ 
$\underline{E}, E_{end}$ & 2500 kWh \\
$\overline{P}_{ESS}$ & 8000 kW \\
$\eta_c$ & 0.95 \\
$\eta_d$ & 1/0.95 \\
$C_r$ & 0.04 \$/kWh \\
$C_{ESS}$ & 0.02 \$/kWh \\ 

$\overline{P}_{dg}$ & 11000 kW \\ 
$\underline{P}_{dg}$ & 1000 kW \\
$RU, RD$ & 3000 kW \\
$SU, SD$ & 4000 kW \\
$C_{dg}$ & 0.65 \$/kWh \\
$DG_{start}$ & 10,000 kWh \\
$DG_{stop}$ & 20,000 kWh \\
$\kappa$ & 0.7 \\

DQN Agent Network & [64, 128, 128, 64] \\
DQN Discount Factor & 0.9 \\
Learning Rate & 0.001 \\
$|A|$ & 40 \\
$w_o$ & 1/1000 \\
$w_b$ & 100 \\

\bottomrule

\end{tabular}
\label{tab:parameters}
\end{table}

\begin{table*}[h]
\caption{\new{The unit cost (\$/kWh) of power import/export with the main grid for different hours of the day.}}
\centering
\fontfamily{ptm}\selectfont
\adjustbox{max width=\textwidth}{
    \begin{tabular}{llllllllllllllllllllll}
    \toprule
    \multicolumn{1}{c}{}      & \multicolumn{21}{c}{Hour of the Day} \\ 
    \multicolumn{1}{l}{} & 00    & 01    & 02    & 03    & 04    & 05    & 06    & 07    & 08    & 09    & 10    & 11-14 & 15    & 16    & 17    & 18    & 19    & 20    & 21    & 22    & 23    \\
    \midrule
    \multicolumn{1}{l|}{\begin{tabular}[c]{@{}l@{}}$C^t_g$\end{tabular}} & 0.144 & 0.151 & 0.142 & 0.130 & 0.117 & 0.116 & 0.121 & 0.136 & 0.138 & 0.144 & 0.198 & 0.144 & 0.150 & 0.210 & 0.197 & 0.240 & 0.321 & 0.330 & 0.315 & 0.244 & 0.260 \\
    \bottomrule
    \end{tabular}
}
\label{tab:parameters2}
\end{table*}


\begin{table}[]
\caption{Average Operation Cost of different approaches (lower the better).}
    \centering
    \fontfamily{ptm}\selectfont
    \begin{tabular}{lll}
    \toprule
    \begin{tabular}[c]{@{}l@{}} Real-Time \\ Operation \end{tabular} & \begin{tabular}[c]{@{}l@{}} Average Hourly \\ Total \\ Operation Cost (\$) \end{tabular} & 
    \begin{tabular}[c]{@{}l@{}} Average \\ Hourly \\ DG Cost (\$) \end{tabular}
    \\ \midrule
    \begin{tabular}[l]{@{}c@{}}MPC Perfect \\ (Benchmark) \end{tabular} & 1809 & 912 \\ 
    DRL & 1816 & 917\\
    MPC Stochastic & 1839 & 943\\ 
    MPC Forecast & 1875 & 977 \\ 
    Rule-Based & 1917 &  1011\\ 
    \bottomrule
    \end{tabular}
    \label{tab:results}
\end{table}

\begin{figure}[!htb]
    \centering
    \includegraphics[scale=0.38]{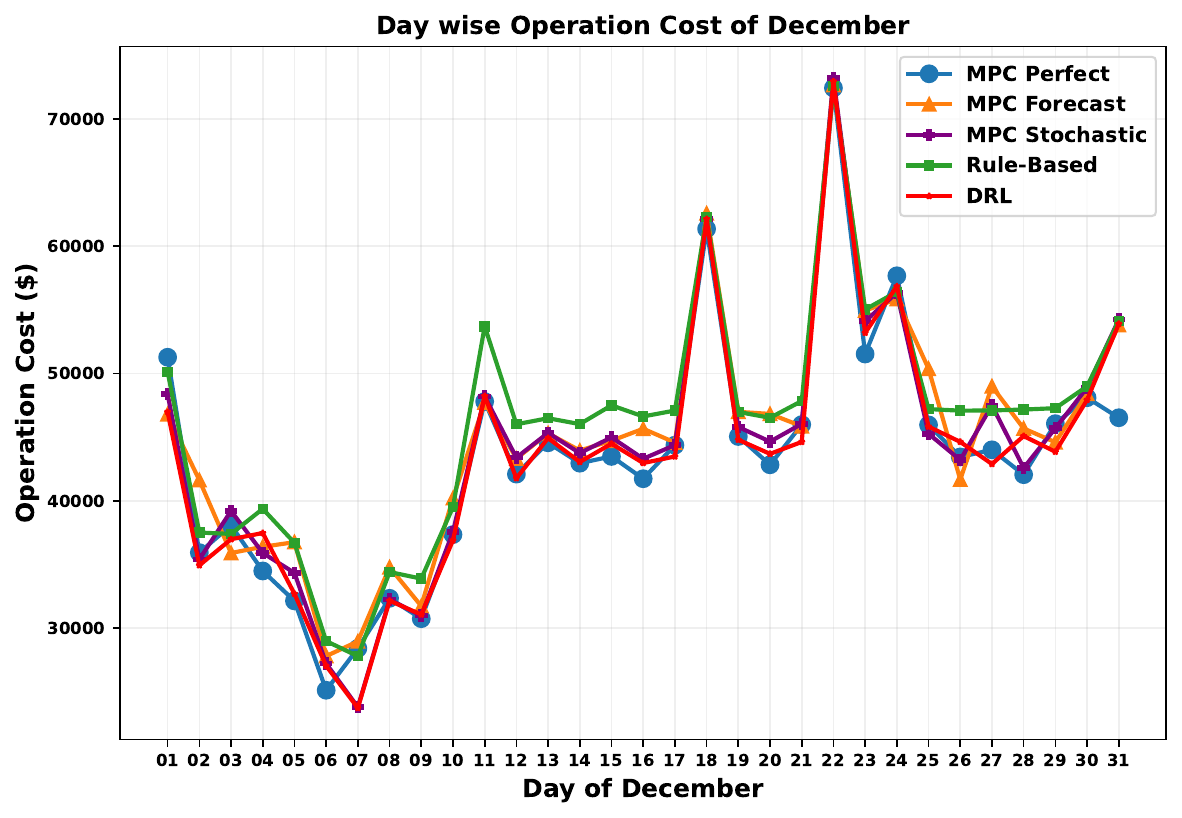}
    \captionsetup{justification=centering}
    \caption{Day wise operation cost in Microgrid.}
    \label{fig:day_wise}
\end{figure}

\subsection{Experiment Results}
Figure \ref{fig:day_wise} shows the daily operation cost of the different algorithms over December. The MPC with a perfect forecast is used to assess the performance of the methods for real-time operation. Table \ref{tab:results} presents the average hourly total operation cost and average hourly DG cost. It is evident from Table \ref{tab:results} that the cost of operating the DG dominates the cost of operating the microgrid - i.e. almost half of the total cost. The results show that the rule-based approach performs the worst among other approaches for real-time operation, as it does not use any additional information such as forecasts or historical data. The performance of MPC Forecast relies heavily on the forecasting techniques, which is challenging in the uncertain SEA context, as shown in Figure \ref{fig:sea_vs_us}. It is apparently outperformed by MPC Stochastic, which does not consider the weather of previous days. This demonstrates that the non-stationary distribution of load demand and solar generation profiles from SEA is better handled by weighting different profiles rather than attending to predict the next steps. Nevertheless, DRL, whose decisions are highly dependent upon the last few weather data, but also learns to consider the distribution of profiles from it, performs the best, which demonstrates its ability to handle SEA weather uncertainties even better than using a fixed profile.

In the following subsections, we discuss the behaviors of the different approaches by analyzing three consecutive days: December $26^{th}$, $27^{th}$, and $28^{th}$, as the performance of each day is dependent on the previous days.

\begin{figure*}[!htb]
\centering
    \subfloat[]{\includegraphics[width=2.23in]{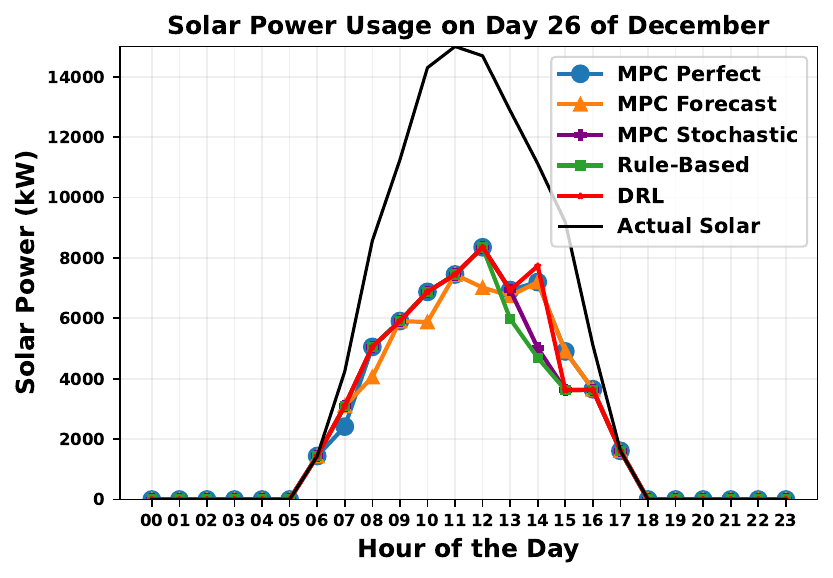}
    \label{fig:day26a}}
    \hfil
    \subfloat[]{\includegraphics[width=2.23in]{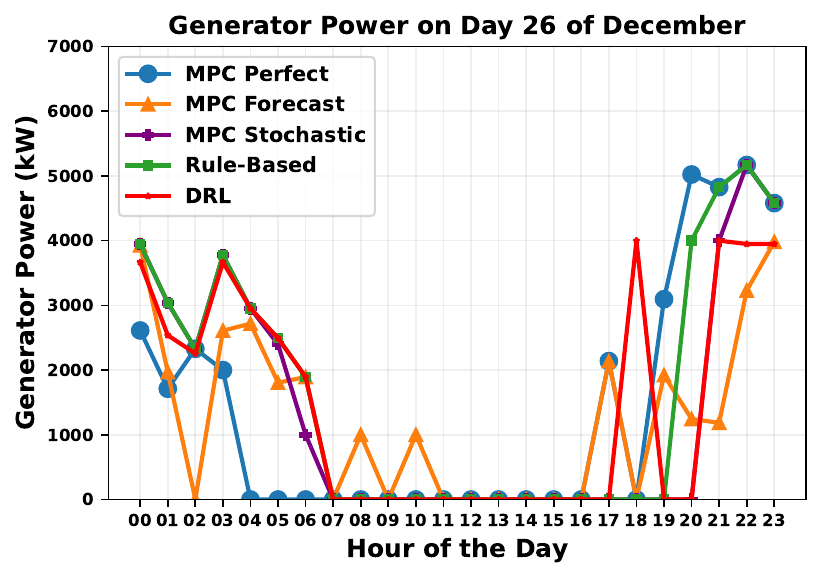}%
    \label{fig:day26b}}
    \hfil
    \subfloat[]{\includegraphics[width=2.23in]{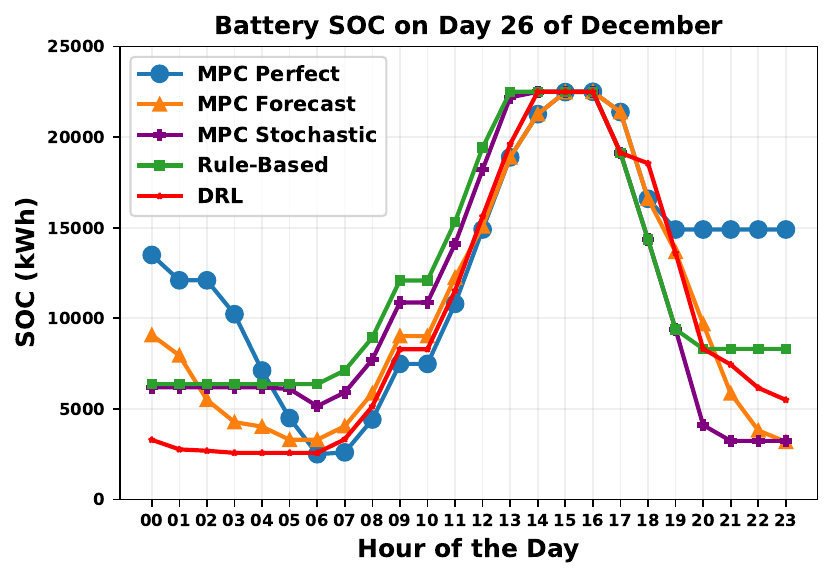}%
    \label{fig:day26c}}
    
\caption{Results for Day 26. Operation Cost for the day: MPC Perfect (\$43452.06), MPC Forecast (\$41671.89), MPC Stochastic (\$43213.77), Rule-Based (\$47084.47) and DRL (\$44635.78).}
\label{fig:day26}
\end{figure*}

\begin{figure*}[!htb]
\centering
    \subfloat[]{\includegraphics[width=2.23in]{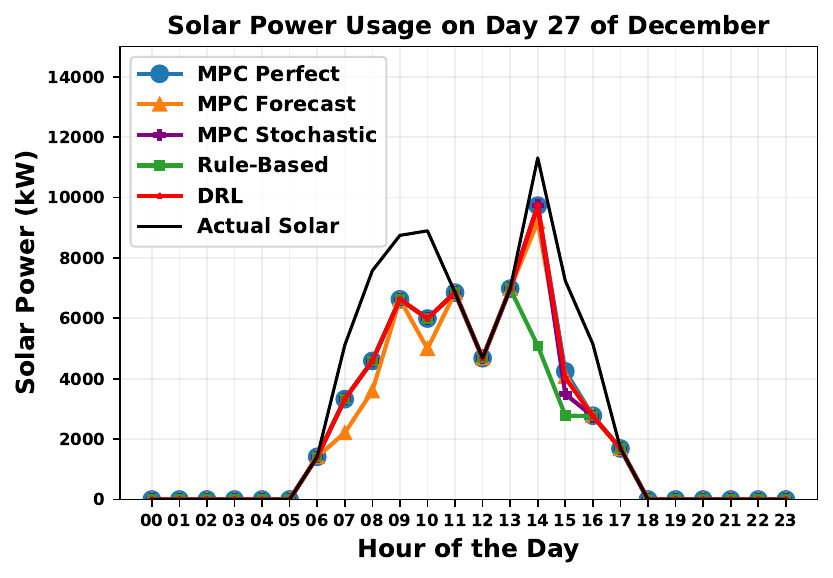}
    \label{fig:day27a}}
    \hfil
    \subfloat[]{\includegraphics[width=2.23in]{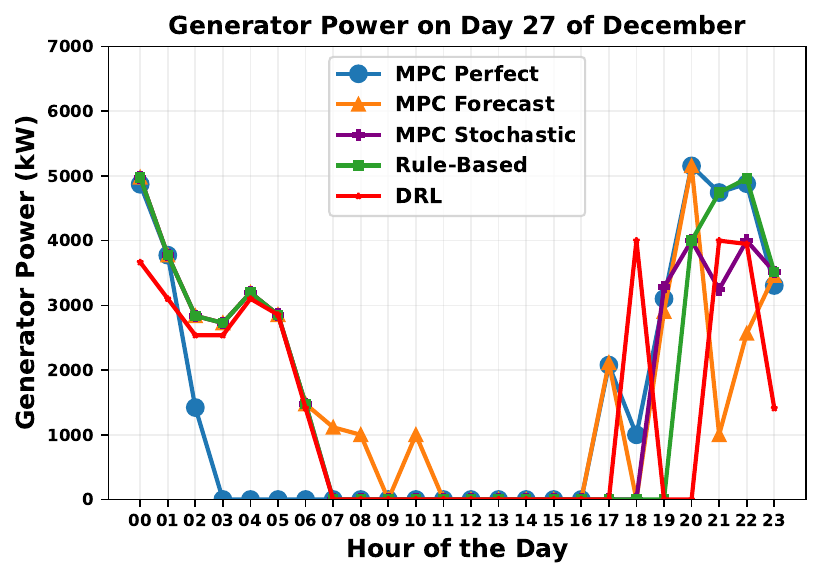}%
    \label{fig:day27b}}
    \hfil
    \subfloat[]{\includegraphics[width=2.23in]{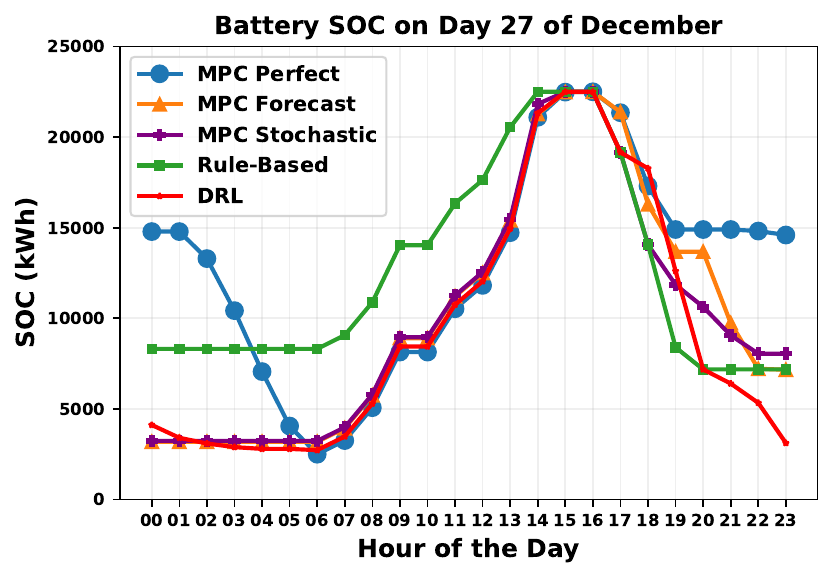}%
    \label{fig:day27c}}
    
\caption{Results for Day 27. Operation Cost for the day: MPC Perfect (\$44009.79), MPC Forecast (\$48990.83), MPC Stochastic (\$47496.56), Rule-Based (\$47103.38) and DRL (\$42896.43).}
\label{fig:day27}
\end{figure*}

\begin{figure*}[!htb]
\centering
    \subfloat[]{\includegraphics[width=2.23in]{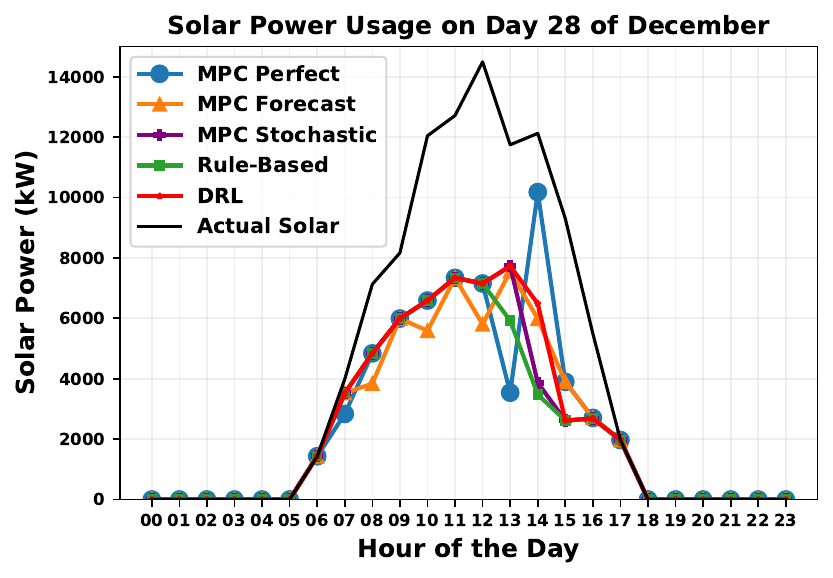}
    \label{fig:day28a}}
    \hfil
    \subfloat[]{\includegraphics[width=2.23in]{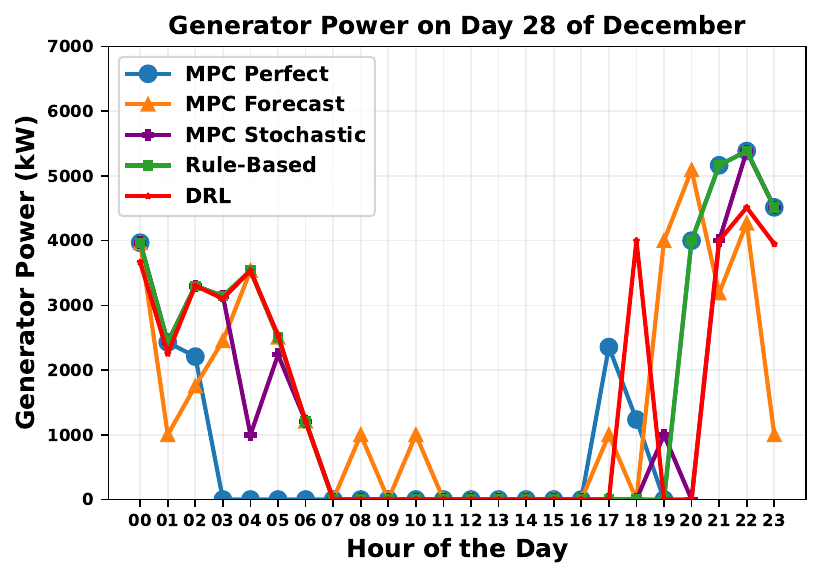}%
    \label{fig:day28b}}
    \hfil
    \subfloat[]{\includegraphics[width=2.23in]{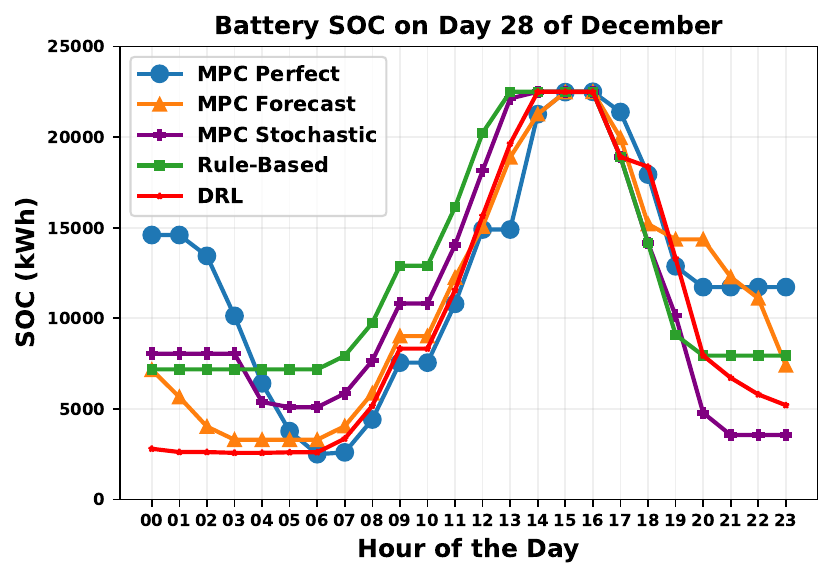}%
    \label{fig:day28c}}
    
\caption{Results for Day 28. Operation cost for the day: MPC Perfect (\$42060.02), MPC Forecast (\$45686.74), MPC Stochastic (\$42591.59), Rule-Based (\$47173.11) and DRL (\$45093.41).}
\label{fig:day28}
\end{figure*}

\begin{figure*}[!htb]
\centering
    
    \subfloat[]{\includegraphics[width=2.23in]{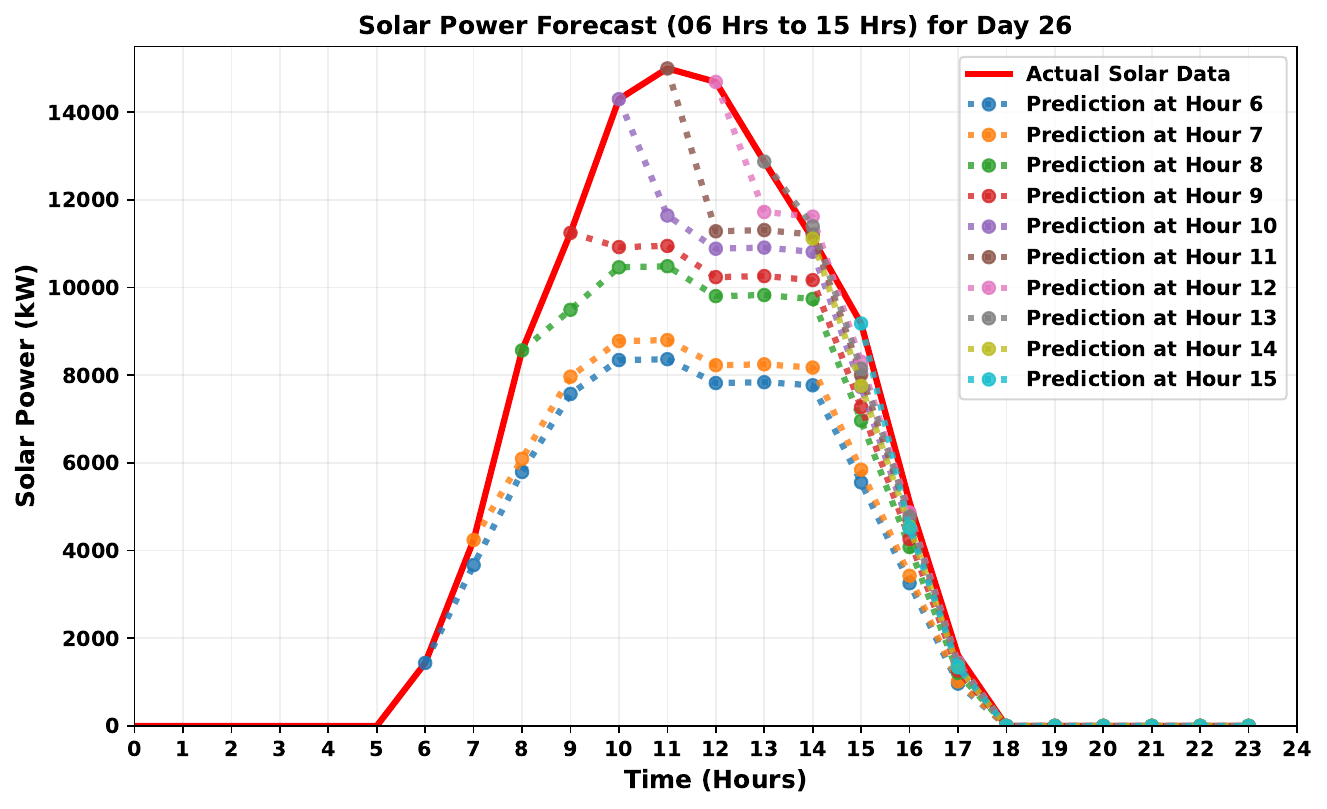}%
    \label{fig:forecast26}}
    \hfil
    \subfloat[]{\includegraphics[width=2.23in]{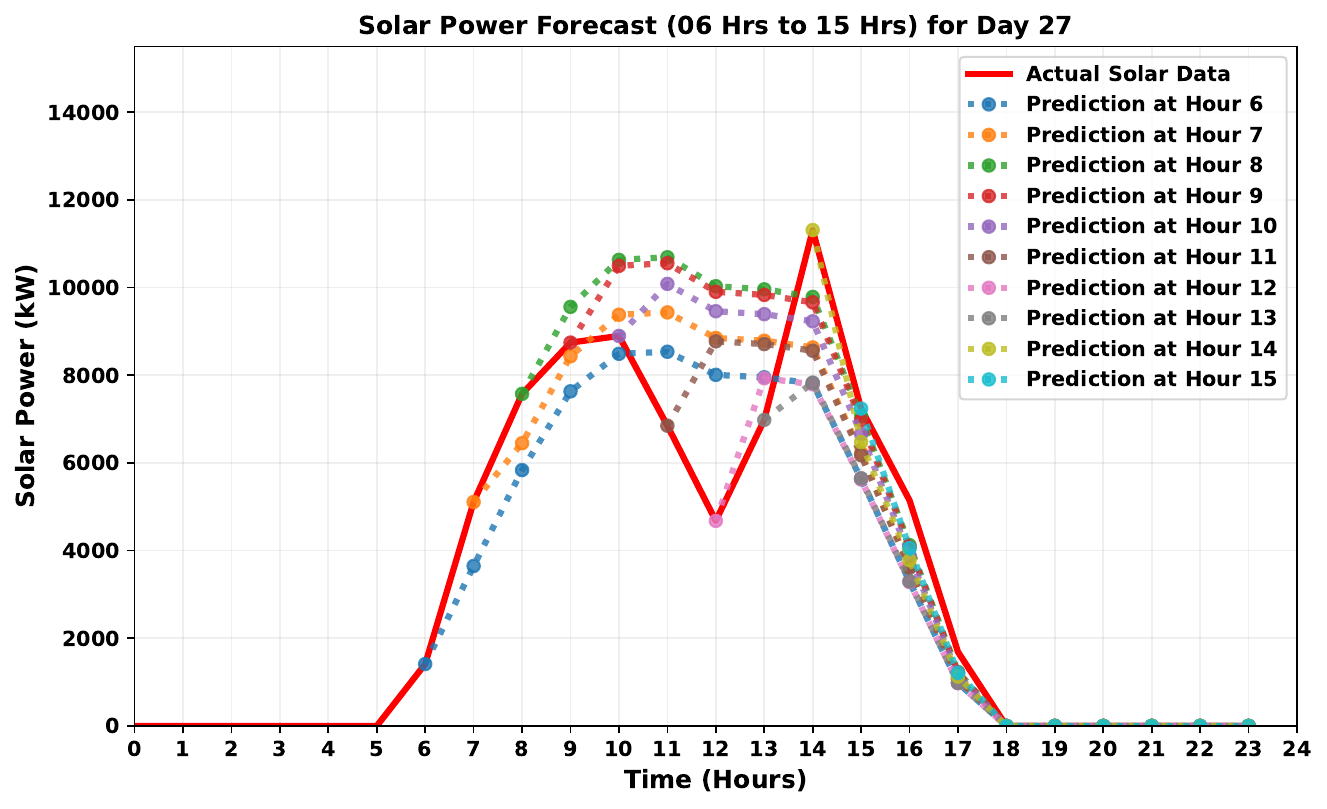}%
    \label{fig:forecast27}}
    \hfil
    \subfloat[]{\includegraphics[width=2.23in]{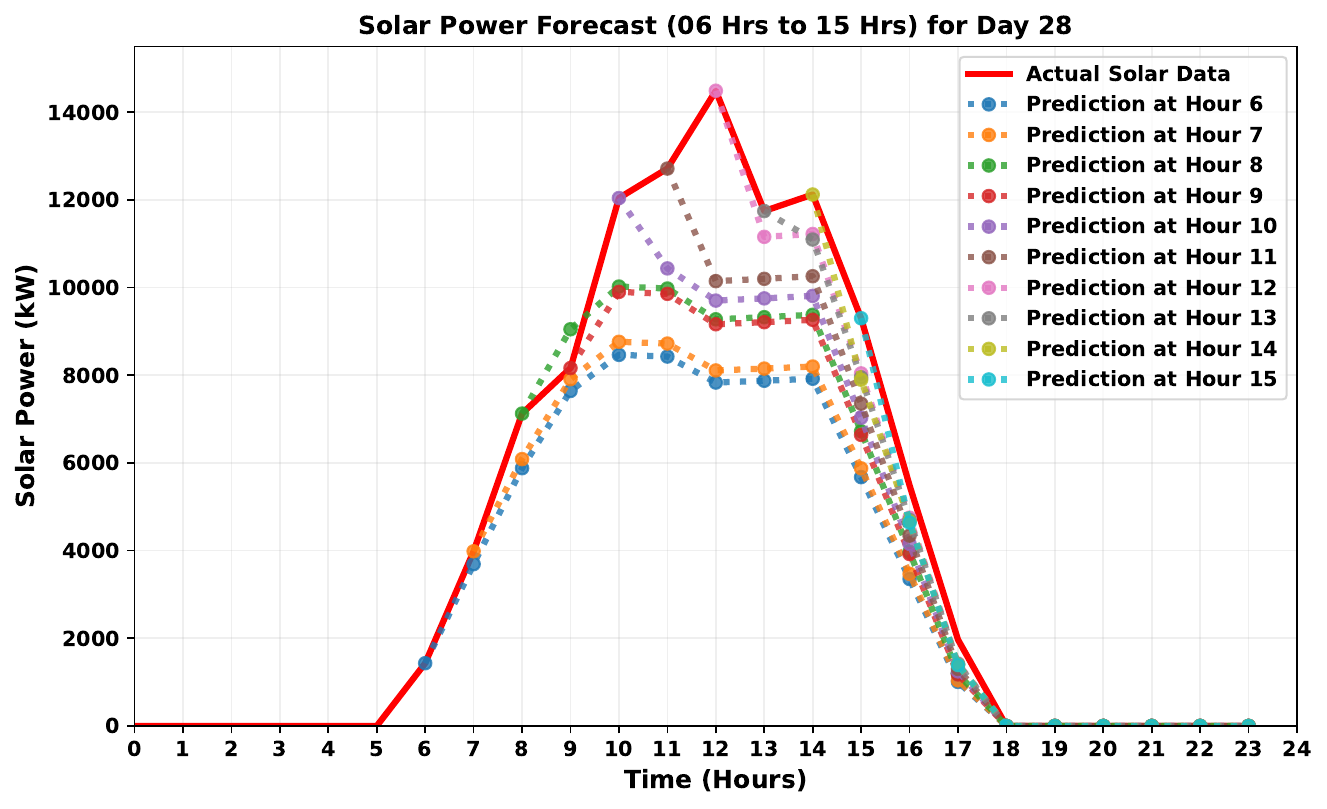}%
    \label{fig:forecast28}}
    
\caption{PV Generation Forecast used in MPC Forecast for a) day 26, b) day 27 and c) day 28.}
\label{fig:forecast}
\end{figure*}

\subsubsection{Behavior of Rule-based}

\begin{table}[]
\caption{Execution Time (seconds) of different approaches. Ordered from low to high.}
    \centering
    \fontfamily{ptm}\selectfont
    \begin{tabular}{ll}
    \toprule
    Real-Time Operation & Average Execution Time (s) \\ \midrule
    Rule-Based & 0.041 \\ 
    DRL & 0.051 \\ 
    MPC Perfect & 0.064  \\ 
    MPC Forecast & 0.118  \\ 
    MPC Stochastic & 0.125 \\ \bottomrule
    
    \end{tabular}
    \label{tab:exe_time}
\end{table}

The performance of the rule-based algorithm is consistently suboptimal.
The main reason is that it does not even take into account the time of day in its operation. For instance, every other approach has information about the sunrise at 6 am, and that they will have PV energy at that time. Thus, all other approaches tend to deplete the ESS SOC to its lowest level at 6 am (true on all three days, though not fully for MPC Stochastic) so that it can be recharged during the day. On the contrary, the rule-based approach keeps a higher level of SOC at 6 am, causing the ESS to reach its full capacity faster than any other approach. This somewhat leads to a waste of solar energy around 2 pm or 3 pm when it has no more capacity to store energy.

As depicted in Table \ref{tab:exe_time}, the computation time for the rule-based algorithm is however the lowest. Despite its suboptimal performance, it is not too far from the other algorithms. This makes the rule-based approach valuable in situations where computation is performed on resource-constrained edge devices, where the use of optimizer toolboxes is challenging.

\subsubsection{Behavior of MPC Forecast}

MPC Forecast depends heavily upon the forecast, which is itself influenced predominantly by recent events. At the end of Day 26 (sunny), the MPC Forecast foresees a sunny next day, hence retaining only a minimal SOC for the ESS \new{(3000kWh)} (the reason why the cost for Day 26 is lowest for MPC Forecast). On the other hand, Day 27 is cloudy, and consequently, by the end of the day, it maintains a higher SOC \new{(7000kWh)}, leading to its highest cost on Day 27.

Although the forecast is influenced by the previous day, it can adapt to the present conditions based on data from the last few hours. For example, consider the forecast at 6 am in Figure \ref{fig:forecast26} for Day 26. The forecast for the day at that time is significantly less than the actual PV generation due to the influence of the previous day. However, after a few hours, around 9 am, the forecast adapts to the current day, closing the gap between the forecasted and the actual PV generation. Similarly, adaptation can be observed when transitioning from a sunny day (Day 26) to a cloudy day (Day 27) in Figure \ref{fig:forecast27} and a cloudy day (Day 27) to a sunny day (Day 28) in Figure \ref{fig:forecast28}.

In geographical regions where the weather is smooth, MPC Forecast can perform well for real-time microgrid operation. This is not the case in SEA: for instance, it can be observed that there was an abrupt decline in PV power generation around 12 PM on Day 27 in Figure \ref{fig:forecast27}, possibly due to an unexpected passing cloud.

\subsubsection{Behavior of MPC Stochastic}

MPC Stochastic has, by construction, a cautious behavior because it considers all five profiles (from very sunny to very cloudy days) to optimize its decision. In particular, it does not use as much ESS power by 6 am as the other approaches (except for Rule-based, which does not take time into account anyway). Hence, it does not have to use DG at 8 or 10 am, which the MPC Forecast needs to do. As a result, it can waste more solar energy than other methods (e.g. on a sunny Day 26 at 2 pm), but overall, the trade-off pays off, as DG energy is the costliest.

Additionally, because of this cautious behavior, on Day 27, a cloudy day, MPC Stochastic has energy stored in the ESS at midnight (00 Hrs). It resulted in lesser DG power consumption during the early morning hours, reducing operation costs on Day 28.
In the context of our SEA dataset, having a fixed prior distribution provides sufficient information to perform well in real-time operations.

\subsubsection{Behavior of DRL}

DRL is performing the most efficiently of all usable approaches and is very close to the absolute optimal MPC Perfect (which cannot be used in practice as it assumes to forecast perfectly, which is impossible):
1816 vs 1809$\$/$hour in average, or $0.3\%$. It shows that a balance between MPC Stochastic and MPC Forecast can be even more efficient than MPC Stochastic. Indeed, DRL takes into account a distribution of profiles, similar to MPC Stochastic, and at the same time, it considers the latest PV generation data to optimize its decision, as MPC Forecast. The learning-based (DRL) approach is trained on historical data and can make bold decisions that help to reduce operating costs. Nevertheless, there are certain days, like Day 28, where DRL couldn't perform well in real-time operation. It is due to the lack of a reasonable sized dataset to learn the nuances of uncertainties in renewable generation.

\begin{figure}[!htb]
    \centering
    \includegraphics[scale=0.5]{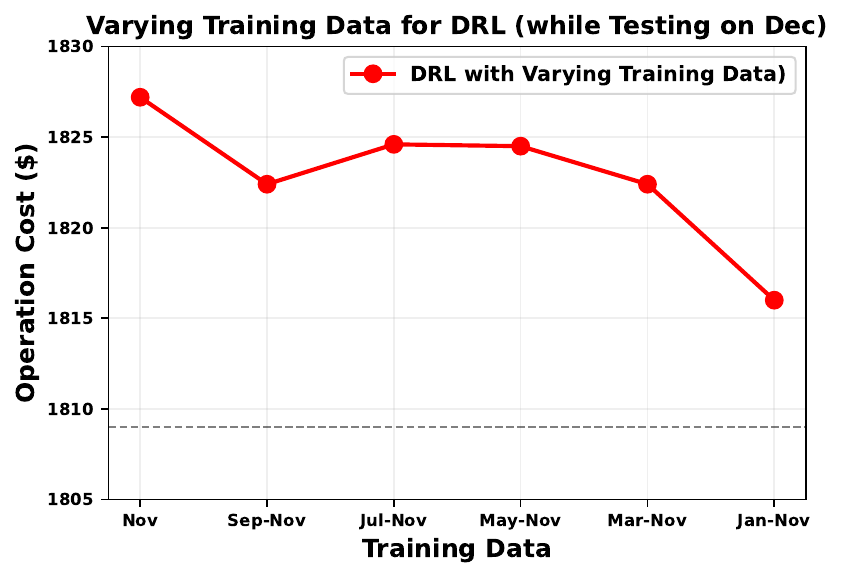}
    \captionsetup{justification=centering}
    \caption{Training DRL by varying the training data size to analyze the effect of the quantity of training data on DRL performance.}
    \label{fig:vary_train_data}
\end{figure}

To demonstrate the impact of training data quantity on the performance of DRL, we gradually increase the amount of training data while evaluating the results of December's data. Initially, we used only November data for training. Then, we keep augmenting the data by including data from September to November and continue until we use the data from January to November. Figure \ref{fig:vary_train_data} illustrates the effect of the amount of training data on DRL's performance. Specifically, increasing the amount of training data improves the performance of DRL as it can learn more subtleties in the data. The variance in operation cost appears to be very low. It can be observed that using data only from November for training results in an average operation cost of 1827 on the test dataset, which is still better than all the other approaches. However, utilizing data from January to November slightly improves the performance to an average operation cost of 1816.

\section{Conclusion} \label{conclusion}
This paper introduced a hierarchical management strategy for power dispatch in a grid-connected microgrid, which involves formulating a day-ahead power exchange commitment with the main grid as SP. Subsequently, power exchange set points of the DG and ESS were determined in real-time operation with hourly granularity while strictly adhering to the grid exchange commitment. We conducted experiments using a real-world dataset to investigate various strategies for handling weather uncertainties in microgrid operation, such as the MPC Forecast, MPC Stochastic, Rule-Based, and learning-based (DRL) approaches. The results showed that the DRL-based strategy outperformed the other approaches in real-time operation with the given dataset, owing to its ability to handle uncertainty. \new{One of the limitations identified for this work thus far is that uncertainties arising from model approximation in the controller have not been taken into account. Indeed, in the implemented simulations, the same model equations for the DG and ESS were considered in the controller phase and in the evaluation of the system performances. In actual deployment, the behavior of those assets cannot be perfectly predicted – e.g., depletion of ESS storage capacity due to its usage, part load operation efficiency of the DG engine, etc. The mitigation of those additional sources of uncertainties would require an online tuning of the system models in the controllers and will be the subject of further work. Furthermore, the real-time operation in this work is based on hourly granularity/schedule, which could be another limitation since the impact of uncertainties tends to increase at finer granularity. Future work may extend to include real-time operations with finer granularity, potentially enhancing the precision and efficiency of power dispatch in microgrids.}

\bibliographystyle{model1-num-names}
\bibliography{mgrid_SEGAN23}

\end{document}